\documentstyle[11pt,fullpage,citesort,epsfig,draft]{article}

\def\Tr{{\rm Tr}}
\newcommand{\ket}[1]{|#1\rangle}
\newcommand{\bra}[1]{\langle#1|}

\def\equalatn{{{}_{{{\displaystyle=}\atop {\displaystyle N}}}}}
\def\ie{{\it i.e.\ }}
\def\m@th{\mathsurround=0pt }
\def\leftrightarrowfill{$\m@th \mathord\leftarrow \mkern-6mu
 \cleaders\hbox{$\mkern-2mu \mathord- \mkern-2mu$}\hfill
 \mkern-6mu \mathord\rightarrow$}
\def\overleftrightarrow#1{\vbox{\ialign{##\crcr
     \leftrightarrowfill\crcr\noalign{\kern-1pt\nointerlineskip}
     $\hfil\displaystyle{#1}\hfil$\crcr}}}
\def\VEV#1{\langle#1\rangle}

\begin{document}
\renewcommand{\thefootnote}{\fnsymbol{footnote}}
\begin{titlepage}
\begin{flushright}
LBNL-49194\\
UCB-PTH-01/43\\
UFIFT-HEP-01-23\\
hep-th/0111280
\end{flushright}
\vskip 2.5cm

\begin{center}
\begin{Large}
{\bf Large $N$ Matrix Mechanics on the Light-Cone
}
\end{Large}

\vskip 2.cm

{\large M. B. Halpern\footnote{E-mail  address: 
{\tt halpern@physics.berkeley.edu}}
 and Charles B. 
Thorn\footnote{E-mail  address: 
{\tt thorn@phys.ufl.edu}}\footnote{Visiting Miller 
Research Professor, on sabbatical leave
from the Department of Physics, University of Florida, Gainesville
FL 32611.} }
\vskip 0.5cm

{\it Department of Physics, University of California,
Berkeley CA 94720}

\vskip0.15cm
and
\vskip0.15cm
{\it Theoretical Physics Group, Lawrence Berkeley National
Laboratory\\ University of California,
Berkeley CA 94720
}

\vskip 1.0cm
\end{center}

\begin{abstract}\noindent
We report a simplification in the large $N$ matrix mechanics of light-cone
matrix field theories. The absence of pure creation or pure annihilation
terms in the Hamiltonian formulation of these theories allows us to find
their reduced large $N$ Hamiltonians as explicit functions of  the generators
of the Cuntz algebra.  
This opens up a free-algebraic playground of new
reduced models -- all  of which exhibit new hidden conserved quantities at
large $N$ and all of  whose  eigenvalue problems are surprisingly simple.
The basic tool we develop for the study of these models is the infinite
dimensional  algebra of all normal-ordered products of Cuntz operators, and
this algebra also leads us to a special number-conserving 
subset of these models, each of which
exhibits an infinite number of new hidden conserved quantities at large $N$.  
\end{abstract}
\vfill
\end{titlepage}

\section{Introduction}
Light-cone quantum field theories 
\cite{diracfront,weinbergfront,susskindgal,fryes,bardakcihlc} 
have been studied since 1949 and
large $N$ methods \cite{thooftlargen,brezinipz} since 1974.  
See Refs.\cite{brodskyppreport,halperns} for more
complete referencing of the two developments.
The idea of combining light-cone quantization with large
$N$ was pursued in \cite{thooftlargen,thornfishnet} in
the context of summing planar Feynman diagrams, and in
\cite{thornfock,dalleyk} in the context of solving the large $N$ energy
eigenvalue problem in the color singlet sector of the light-cone Fock space.
Here we study a different combination of the two developments--which
leads to an algebraic reformulation of the light-cone
theory at large $N$.

In this paper we will focus on large $N$ phase space methods 
\cite{jevickis,bardakci81,bardakcih81,halpern812} and
in particular on the development known as {\it large $N$ matrix mechanics} 
\cite{halpern812,halperns81,halpern82}, \cite{halperns},
\cite{halperns99,schwartz}. 
In large $N$ matrix mechanics,  the Hilbert space is spanned by
the singlet ground state and the dominant
adjoint eigenstates which saturate the traces of the theory
at large N, and the output is the so-called {\it reduced formulation} for both
large $N$ action \cite{halperns99} and Hamiltonian \cite{halperns,halperns99} 
theories  in terms of {\it free
algebras}.  The simplest free algebra found at large $N$ 
is the {\it Cuntz algebra}, whose history in
mathematics and physics can be found in Ref.~\cite{halperns}.  
In fact, a broad
variety of free algebras has been observed 
\cite{halpern812}, \cite{halperns,halperns99}
in large $N$ matrix
mechanics, including the Cuntz algebra,  the symmetric Cuntz algebra, the
interacting symmetric Cuntz algebras and infinite dimensional free
algebras, as well as fermionic versions and Cuntz superalgebras.  Moreover,
the algebraic formulation  has led to the discovery of new hidden conserved
quantities \cite{halperns}  which appear only at large N, as well as various
free-algebraic forms of the master field (see also Ref.\cite{gopakumarg}), 
and a numerical approach \cite{schwartz}  
to computations in many- matrix models at large N.

But there has always been an enigma, called the opacity phenomenon
in Ref.\cite{halperns},  in the
large $N$ matrix mechanics of equal-time quantized Hamiltonian formulations:
Except for  oscillators \cite{halperns} and one matrix models 
\cite{halperns81}, the reduced free-algebraic
Hamiltonian of a given large $N$ system is difficult to obtain in closed
form.

We report  here an important simplification in the 
large N matrix mechanics of  discretized light-cone matrix quantum 
field theories, including light cone models which are both local and 
non-local in $x^-$. 
The non-local models, including for example the number-conserving
truncations of local models, can be considered as quantum-mechanical 
models in their own right. Because of the well-known 
absence of pure annihilation and pure creation terms in any
light-cone Hamiltonian $H_{\cdot}= {P.}^-$ we are able to obtain 
{\it the explicit form
of the large $N$ reduced Hamiltonian}  entirely in terms of Cuntz operators.

This opens up  a free-algebraic playground of new
reduced models, all of which exhibit an unexpectedly simple algebraic
structure. In particular, each of these models has {\it new hidden conserved
quantities} at large $N$ and the  free-algebraic eigenvalue problems of
the reduced Hamiltonians are surprisingly simple.

The basic tool which we develop for the study of these models is the
infinite dimensional algebra of all normal-ordered Cuntz operators, and
this algebra also leads us to a special subset of the number-conserving 
models which
apparently becomes {\it integrable} at large N. In particular,  each of these
models exhibits an infinite number of new hidden conserved quantities at
large N, and moreover the eigenvalue problems of these models  exhibit
oscillator-like spectra.

\section{Light-cone Matrix Quantum Field Theories}
We begin with some general remarks about light-cone
quantization of matrix field theories, deferring consideration of large $N$
matrix mechanics to Section 3.
\subsection{The Light-Cone Field}
The matrix field theory of most obvious physical interest is
non-abelian gauge theory. But to simply
illustrate the main ideas, we shall focus here on a
much simpler example, the hermitian scalar $N\times N$
matrix field $\phi_{rs}$ 
in two space-time dimensions. We define light-front coordinates 
$x^\pm=(x^0\pm x^3)/\sqrt2$, choosing $x^+$ as time, 
and referring the $x^-$ coordinate to its conjugate momentum
labelled by $p^+>0$. 
The field $\phi_{rs}$ then has the expansion
\begin{eqnarray}
\phi_{rs}(x^-,x^+)&=&\int_{0^+}^\infty {dp^+\over\sqrt{4\pi p^+}}
\left(a(p^+,x^+)_{rs}e^{-ix^-p^+} +a^\dagger(p^+,x^+)_{rs}e^{+ix^-p^+}
\right)
\end{eqnarray}
with $r,s=1,\ldots, N$. 
The quantum nature of the field is fixed by imposing the
equal time (equal $x^+$) commutation relations
\begin{eqnarray}
\left[a(p^+)_{{rs}},a^\dagger(q^+)_{{tu}}\right]&=& 
\,\delta_{{st}}\delta_{ru}\delta(p^+-q^+) \nonumber\\
\left[a(p^+)_{{rs}},a(q^+)_{{tu}}\right] &=& 
\left[a^\dagger(p^+)_{{rs}},a^\dagger(q^+)_{{tu}}\right]=\,0.
\end{eqnarray}
In the following we shall suppress the $x^+$ argument in expressions
at equal time.
A simple consequence of these commutation relations is
the well-known equal time light-cone field commutation rule
\begin{eqnarray}
\left[\phi_{{rs}}(x^-),\phi_{{tu}}(y^-)\right]&=&
\,{i\over4}\delta_{{st}}\delta_{ru}\epsilon(y^--x^-)
\end{eqnarray}
where $\epsilon(x)=x/|x|$, the sign of $x$. In particular,
the fields at different spatial points
do {\it not} commute, even at equal time.

In our discussion we find it convenient 
to deal with discretely labelled operators so
we discretize $p^+=m\delta$, $m=1,2,\ldots$. 
Putting $a(m\delta)
\to a_{m}/\sqrt{\delta}$, the commutation relations then read
\begin{eqnarray}
\left[(a_m)_{{rs}},(a^\dagger_n)_{{tu}}\right]&=& 
\,\delta_{mn}\delta_{{st}}\delta_{ru} \nonumber\\
\left[(a_m)_{{rs}},(a_n)_{{tu}}\right] &=& 
\left[(a^\dagger_m)_{{rs}},(a^\dagger_n)_{{tu}}\right]=\,0
\label{urcr}\\
(a_m)_{rs}\ket{0.}&=&\bra{.0}(a^\dagger_m)_{rs}=0
\end{eqnarray}
where $\ket{0.}$ is the Fock vacuum, and the expansion of the field reads
\begin{eqnarray}
\phi_{rs}(x^-)&=&\sum_{m=1}^\infty {1\over\sqrt{4\pi m}}
\left((a_{m})_{rs}e^{-ix^-m\delta} +(a^\dagger_{m})_{rs} e^{+ix^-m\delta}
\right)\\
&\to&\sum_{m=1}^K {1\over\sqrt{4\pi m}}
\left((a_{m})_{rs}e^{-ix^-m\delta} +(a^\dagger_{m})_{rs} e^{+ix^-m\delta}
\right)
\label{8}
\end{eqnarray}
where $K$ is a large integer serving as a cutoff on the high $p^{+}$
modes. With $m$ an integer, we should restrict $-\pi\leq x^{-}\delta\leq\pi$.
The expansion (\ref{8}) can be inverted 
\begin{eqnarray}
(a_{m})_{rs}=\delta\sqrt{m\over\pi}\int_{-\pi/\delta}^{\pi/\delta}
e^{+ix^-m\delta}\phi_{rs}(x^-).
\label{9}
\end{eqnarray}
We also need the  contraction factor
\begin{eqnarray}
\int{dp^{+}\over4\pi p^{+}}\to\sum_{m=1}^{\infty}{1\over4\pi m}
\to L\equiv\sum_{m=1}^{K}{1\over4\pi m}
\end{eqnarray}
which arises in the process of normal ordering.
\subsection{Partial Normal Ordering}
For the purposes of this paper the most important feature of
light-cone coordinates is that the creation operators $a^\dagger(p^+)$
always create particles of {\it positive} $p^+$. As seen in Eq.~(1),
we will assume
that $p^+=0$ is excluded, either by a small low momentum cutoff
$\epsilon$, or in the context of discretized $p^+=m\delta$, by the exclusion
of $m=0$. Only annihilation operators carry negative $p^+$. Therefore
the only $p^+$ conserving terms that can appear in the Hamiltonian
contain both creation and annihilation operators, \ie there can
be no terms $\Tr\ a^n$ or $\Tr\ a^{\dagger n}$, $n>0$. As we shall see,
this feature is
responsible for a dramatic simplification in the free-algebraic
description of the large $N$ limit.

Since the creation and annihilation operators are $N\times N$
matrices, as well as operators in Hilbert space, the process
of normal ordering each term in the Hamiltonian can lead to
awkward color descriptions. For example, the normal-ordered
form of $\Tr a^\dagger a a^\dagger a$ would be
\begin{eqnarray}
\Tr :a^\dagger a a^\dagger a: =a^\dagger_{rs}a^\dagger_{tu}a_{st}a_{ur}.
\end{eqnarray}
Here there is a clash between the ordering for matrix multiplication and
the ordering for Hilbert space operator multiplication on the r.h.s.
Thus we refrain from completely normal ordering such terms. 

But we can
{\it partially normal order} the operators in a trace,
so there is always at least one
annihilation operator on the extreme right or at least one
creation operator on the extreme left without disturbing the coincidence of
matrix ordering with quantum mechanical operator
ordering. For example, a partially
normal-ordered form of $\Tr a_1a_2^\dagger a_3 a_4^\dagger$
would be $\Tr a_4^\dagger a_1 a_2^\dagger a_3$. Such
reorderings preserve the cyclic ordering of the operators
within each trace. Note that 
in general a partially normal-ordered form is not unique,
for example $\Tr a_4^\dagger a_1 a_2^\dagger a_3\neq
\Tr a_2^\dagger a_3 a_4^\dagger a_1$ although the
operators are each in the same cyclic order.
We shall see in Secs. 4 and 5, however, that the non-uniqueness
disappears for $N\to\infty$.
Of course, in field theory the Hamiltonian is typically presented
in a (local) form which is not even partially normal-ordered. But
by exploiting the known commutation relations, we can always
rearrange the operators in the partially normal-ordered form
just described. Note that if there is at least one creation and
one annihilation operator in a trace, partial normal ordering
will lead to a generic structure
\begin{eqnarray}
\sum_{m,n}\Tr (a^\dagger_mA_{mn}a_n) 
\end{eqnarray}
where $A$ is any matrix function of $a_p, a^\dagger_q$, not
necessarily itself normal-ordered.

As a  concrete example we consider the $U(N)$-invariant light-cone
Hamiltonian   
\begin{eqnarray}
H.&=&P^-_{\cdot}=
{\mu^{2}\over2}\int dx^{-}\Tr\phi^{2}+{\lambda\over4N}\int dx^{-}
\Tr\phi^{4}
\label{quarticscalarham}
\end{eqnarray}
of a local two dimensional scalar field theory. Then the commutation
relations (\ref{urcr}) are used to obtain the partially
normal-ordered form
\begin{eqnarray}
H.&=&E_{0}+\sum_{m,n=1}^{K}\Tr a^{\dagger}_{m}A_{mn}a_{n}
+\sum_{m,n=1}^{K}\Tr(a^{\dagger}_{m})B_{mn}\Tr(a_{n}).
\label{quarticscalarham2}
\end{eqnarray}
The explicit forms of the quantities appearing here are
\begin{eqnarray}
E_{0}&=&[L^{2}(N^{2}+1)\lambda/4 
+ N^{2}L/2]{2\pi\over\delta},\qquad
B_{mn}=\delta_{mn}{\lambda L\over N}{1\over2m\delta}\\
A_{mn}&=&C_{mn}+D_{mnm_{2}m_{3}}
a^{\dagger}_{m_{2}}a^{\dagger}_{m_{3}}+E_{mnm_{2}m_{3}}
a_{m_{2}}a_{m_{3}}+F_{mnm_{2}m_{3}}\left(
a^{\dagger}_{m_{2}}a_{m_{3}}
+{1\over2}a_{m_{3}}a^{\dagger}_{m_{2}}\right)\\
C_{mn}&=&{\mu^{2}\over2m\delta}\left(1+{3\lambda 
L\over2\mu^{2}}\right)\delta_{mn}, \qquad
D_{mnpq}={\lambda\over8\pi \delta N}
{1\over\sqrt{mpqn}}\delta_{m+p+q,n}\\
E_{mnpq}&=&{\lambda\over8\pi \delta N}
{1\over\sqrt{mpqn}}\delta_{m,p+q+n}, \qquad
F_{mnpq}={\lambda\over8\pi \delta N}
{1\over\sqrt{mpqn}}\delta_{m+p,q+n}
\end{eqnarray}
where repeated indices are always summed.
Since our Hamiltonian is now in partially normal-ordered form,
the Fock vacuum $\ket{0.}$ is manifestly an energy eigenstate with
$$ H.\ket{0.}=E_0\ket{0.}.$$
In the above computation we have retained all the commutator terms arising
from reordering the $a$'s and $a^\dagger$'s 
of the original Hamiltonian. Because our starting
Hamiltonian was at most quartic in the fields the only such
terms are $c$-numbers and quadratic. Furthermore, because of
our treatment of $p^+$, there are never reordering terms with
only annihilation or only creation operators. However the
reordering does alter the original color structure in the sense
that we started with only a single trace and ended up with
reordering terms involving two traces. Partial normal
ordering of the general light-cone Hamiltonian is discussed
in Sec. 5.
\section{ Reduced momentum and reduced number operator}
In this section we begin our study of the large $N$ matrix
mechanics \cite{halpern812,halpern82,halperns81}, \cite{halperns},
\cite{halperns99,schwartz} 
of light-cone field theory, drawing heavily on the methods and results
of Ref.\cite{halperns}

We start with the simpler problems of the momentum operator
$P.\equiv P_{\cdot}^+$ and the number operator ${\hat N.}$
\begin{eqnarray}
P.= P_{\cdot}^\dagger\equiv\sum_mmTr a^\dagger_ma_m,
  &&\qquad{\hat N}.={\hat N.}^\dagger\equiv\sum_m Tr a^\dagger_ma_m\\
\left[P. ,\left({a_m\over\sqrt{N}}\right)_{rs}\right]=-m\left({a_m\over\sqrt{N}}\right)_{rs},
 &&\qquad\left[P., \left({a^\dagger_m\over\sqrt{N}}\right)_{rs}\right]
=m\left({a^\dagger_m\over\sqrt{N}}\right)_{rs}\label{1*}\\
\left[{\hat N}. ,\left({a_m\over\sqrt{N}}\right)_{rs}\right]=-\left({a_m\over\sqrt{N}}\right)_{rs},
 &&\qquad\left[{\hat N}., \left({a^\dagger_m\over\sqrt{N}}\right)_{rs}\right]
=\left({a^\dagger_m\over\sqrt{N}}\right)_{rs}\label{2*}\\
P.\ket{0.}={\hat N}.\ket{0.}=\bra{.0}P.&=&\bra{.0}{\hat N}.=0
\end{eqnarray}
which are two examples of so-called {\it trace class operators} 
in the unreduced
theory. In large $N$ matrix mechanics one then takes matrix elements of the
commutation relations above, using only the ground state $\ket{0.}$ 
and the dominant time-independent
adjoint eigenstates $\ket{rs, A}$ 
which saturate the traces of the theory at large N. The
large $N$ completeness relation for these states reads ($t=x^+$)
\begin{eqnarray}
1.\equalatn\ket{0.}\bra{.0}+\sum_{rs, A}\ket{rs, A}\bra{rs, A},
\qquad {d\over dt}\ket{0.}={d\over dt}\ket{rs, A}=0
\end{eqnarray}
because  large $N$ factorization guarantees that the  singlet ground
state will be sufficient to saturate the singlet channels of the traces.

    Next one introduces reduced matrix elements\footnote{Reduced 
matrix elements were introduced by K.Bardakci in Ref.\cite{bardakci81}.}, 
for example
\begin{eqnarray}
\bra{pq, A}\left({\rho_1(t_1)\over\sqrt{N}}\cdots
{\rho_n(t_n)\over\sqrt{N}}\right)_{rs}\ket{st, B}&=&P_{qp,rt}
\bra{A}{\rho_1}(t_1)\cdots
{\rho_n}(t_n)\ket{ B}\label{wereduction1}\\
\bra{.0}\Tr\left({\rho_1(t_1)\over\sqrt{N}}\cdots
{\rho_n(t_n)\over\sqrt{N}}\right)\ket{0.}&=&N
\bra{0}{\rho_1}(t_1)\cdots
{\rho_n}(t_n)\ket{0}\label{wereduction2}\\
P_{sr,pq}=\delta_{rp}\delta_{sq}-{1\over N}\delta_{sr}\delta_{pq},
\qquad \rho &=& a\quad {\rm or}\quad a^\dagger
\end{eqnarray}
where $t_i=x^+_i$ and   $P$ is a Clebsch-Gordan coefficient. 
Matrix products $(\rho_1\cdots\rho_n)_{rs}$ are called
{\it densities} in the unreduced theory. These
``Wigner-Eckart'' relations  then define the reduced operators  $a_m$ and
$a_m^\dagger$ as well as the reduced  ground state $\ket{0}$ and  the reduced
adjoint eigenstates $\ket{A}$. The reduced eigenstates satisfy the 
reduced completeness relation
\begin{eqnarray}
1=\ket{0}\bra{0}+\sum_A\ket{ A}\bra{ A},\qquad 
{d\over dt}\ket{0}={d\over dt}\ket{A}=0
\label{reducedcomplete}
\end{eqnarray}
at large $N$.

The reduced forms of Eqs. (\ref{1*}) and (\ref{2*}) are
\begin{eqnarray}
\left[P ,{a_m}\right]=-m{a_m},
 &&\qquad\left[P, a^\dagger_m\right]
=ma^\dagger_m\label{3*}\\
\left[{\hat N} ,a_m\right]=-a_m,
 &&\qquad\left[{\hat N}, a^\dagger_m\right]
=a^\dagger_m .\label{4*}
\end{eqnarray}
Here $P$ and ${\hat N}$ are the reduced 
momentum operator and the reduced number
operator respectively, and the reduced operators $a_m$ and $a_m^\dagger$ 
satisfy the {\it Cuntz algebra}
\begin{eqnarray}
a_ma^\dagger_n&=&\delta_{mn},\qquad \sum_ma^\dagger_ma_m=1-\ket{0}\bra{0}
\label{cuntzalg}\\
&&a_m\ket{0}=\bra{0}a^\dagger_m=0 
\label{groundcond}
\end{eqnarray}
where $\ket{0}$ is the reduced  ground state of the system. 
The Cuntz algebra is
a simple example of a so-called free algebra, defined by products
instead of commutators, and this free algebra is known to describe classical or
Boltzmann statistics, another classical aspect of the
large $N$ limit. 

In fact the Cuntz algebra is a subalgebra of the
so-called {\it symmetric Cuntz algebra} \cite{halperns} 
which includes the following further
operators and relations
\begin{eqnarray}
{\tilde a}_m{\tilde a}^\dagger_n&=&\delta_{mn},
\qquad \sum_m{\tilde a}^\dagger_m{\tilde a}_m=1-\ket{0}\bra{0}
\label{tildeops1}\\
\left[ {\tilde a}_m, a^\dagger_n\right]&=&\left[a_m,{\tilde a}^\dagger_n\right]
=\delta_{mn}\ket{0}\bra{0}, \qquad [a_m,{\tilde a}_n]=[a^\dagger_m,
{\tilde a}^\dagger_n]=0 
\label{tildeops2}\\
{\tilde a}_m\ket{0}&=&\bra{0}{\tilde a}^\dagger_m=0,\qquad
{\tilde a}^\dagger_m\ket{0}={a}^\dagger_m\ket{0},\qquad
\bra{0}{\tilde a}_m=\bra{0}{a}_m
\label{tildeops}
\end{eqnarray}
Moreover, the tilde operators ${\tilde a}_m$, ${\tilde a}^\dagger_m$
 are important in
deriving the Cuntz algebra (\ref{cuntzalg}) from the unreduced theory:  
The reduction procedure (see Eq.~(2.29) of Ref.~\cite{halperns})
gives only the relations in Eq. (\ref{tildeops2}) as
a consequence of the unreduced commutators (\ref{urcr}), and a completeness
argument (see Eq.~(3.7) of 
Ref.~\cite{halperns}) on the general basis state is necessary to 
obtain the Cuntz algebra from (\ref{tildeops2}) 
and the ground state conditions. Beyond 
this application, the tilde operators are not used in this paper.

The reduction procedure does not give us directly the composite
structure of reduced trace class operators such as P, $\hat N$ or the reduced
Hamiltonian H. This is called the opacity phenomenon in Ref.\cite{halperns}. 
But one
can in principle find the composite structure of such operators by solving
reduced commutation relations such as  Eqs.(\ref{3*}) and (\ref{4*}).

   In fact, the reduced commutators  (\ref{3*}) and (\ref{4*}) 
are so simple that standard methods \cite{halperns} suffice to determine 
the form of the reduced operators $P$ and ${\hat N}$. To express these
solutions succinctly, it is convenient to introduce a standard tool called
{\it word notation}:
\begin{eqnarray}
a_w&\equiv& a_{m_1}\cdots a_{m_n}, \qquad w=m_1\cdots m_n\\
a_w^{\phantom{w}\dagger}&\equiv& a^\dagger_{m_n}
\cdots a^\dagger_{m_1}\label{aword}\\
\left[ w \right] &\equiv& n, \qquad \{w\}\equiv\sum_{i=1}^n m_i .
\label{lengthw}
\end{eqnarray}
Here $w$ is an arbitrary word composed of  some number of  ordered letters,
which label the lattice sites in this case. 
In (\ref{lengthw}) we have also defined
the length $[w]$  of the word $w$ and its weight $\{w\}$, 
which is the sum of its
letters.  

   Then the reduced momentum and reduced number operators can  be
written as
\begin{eqnarray}
P&=&P^\dagger=\sum_w a_w^{\phantom{w}\dagger}
\left(\sum_m ma_m^\dagger a_m\right) a_w\nonumber\\
&=&\sum_m ma_m^\dagger a_m 
+ \sum_p a^\dagger_p\left(\sum_m ma_m^\dagger a_m\right)a_p 
+ \cdots\label{reducedpee}\\
{\hat N}&=&{\hat N}^\dagger
=\sum_w a_w^{\phantom{w}\dagger}
\left(\sum_m a_m^\dagger a_m\right) a_w\nonumber\\
&=&\sum_m a_m^\dagger a_m 
+ \sum_p a^\dagger_p\left(\sum_ma_m^\dagger a_m\right)a_p + \cdots\\
P\ket{0}&=&{\hat N}\ket{0}=\bra{0}P=\bra{0}{\hat N}=0
\end{eqnarray}
where $\sum_w$ means the sum over all words $w$, including the
trivial word $w=0$.    To check that these
operators satisfy the commutation relations (\ref{3*}) and (\ref{4*}), 
write them out,
e.g. $[P, a_m]= Pa_m - a_mP$, and use the Cuntz algebra (\ref{cuntzalg}) 
to  cancel all but
one of the terms in the difference. 
Later we shall discuss more sophisticated ways to handle
operators of this type.

\section{Reduced Hamiltonian}
The reduced momentum and number operators 
discussed above are very simple examples,
but the opacity phenomenon has been particularly vexing in the case of 
models described by the
equal-time quantized Hamiltonian
\begin{eqnarray}
H.=\Tr\left({1\over2}\pi^m\pi^m+NV\left({\phi\over\sqrt{N}}\right)\right).
\end{eqnarray}
For such systems one knows the reduced equations of motion
\cite{halpern812,halperns}
\begin{eqnarray}
{\dot\phi}_m=i[H,\phi_m]=\pi_m,\qquad
{\dot\pi}_m=i[H,\pi_m]=-V_m(\phi)\\
\rho(t)=e^{iHt}\rho(0)e^{-iHt},\qquad \rho=\phi_m\quad{\rm or}\quad\pi_m
\end{eqnarray}
where $H$ is the reduced Hamiltonian, and one knows the interacting
Cuntz algebras \cite{halperns,halperns99}. 
But beyond the case of matrix oscillators \cite{halperns} 
and the general one-matrix
model \cite{halperns81}, 
the explicit form  of  the reduced Hamiltonian $H$ of such systems is
not known  (an implicit construction of $H$ is given 
in Subsec.~4.6 of Ref.~\cite{halperns}).

For quantized matrix theories of the
{\it light-cone type}, i.e. with no pure creation or
pure annihilation terms in $H.$, we shall see that we can in fact
find the explicit form of the reduced Hamiltonian.

As an example we begin with a general  
unreduced quartic Hamiltonian of the  light-cone type
\begin{eqnarray}
&&\hskip-2in
H.-E_0=N^2\left\{{1\over N}\Tr\left(\sum_{m,n}{a^\dagger_m\over\sqrt{N}}
A_{mn}\left({a^\dagger\over\sqrt{N}},{a\over\sqrt{N}}\right){a_n\over\sqrt{N}}
\right)+\left({1\over N}\Tr{a^\dagger_m\over\sqrt{N}}\right)
B_{mn}\left({1\over N}\Tr{a_n\over\sqrt{N}}\right)\right\}
\label{unreducedh}\\
A_{mn}\left({a^\dagger\over\sqrt{N}},{a\over\sqrt{N}}\right)_{rs}
&=&\left(C_{mn}+{C_{mnp}\over\sqrt{N}}a^\dagger_p+
{D_{mnp}\over\sqrt{N}}a_p+{C_{mnpq}\over{N}}a^\dagger_p a^\dagger_q
\right.\nonumber\\
&&\hskip1cm\left.+{D_{mnpq}\over{N}}a^\dagger_p a_q
+{E_{mnpq}\over{N}}a_p a^\dagger_q
+{F_{mnpq}\over{N}}a_p a_q\right)_{rs}\\
C_{mn}&=&C^*_{nm},\qquad
D^*_{mnp}=C_{nmp},\qquad
C^*_{mnpq}=F_{nmqp}\nonumber\\
D^*_{mnpq}&=&D_{nmqp},\qquad
E^*_{mnpq}=E_{nmqp},\qquad
B^*_{mn}=B_{nm}.
\label{reality}
\end{eqnarray}
Here the quantities $B,C,D,E,F$ are constants 
and the relations in (\ref{reality})
guarantee $A^\dagger_{mn}=A_{nm}$ and hence $H^\dagger_{\cdot}=H.$.
In what follows, we often refer to the form $A_{mn}$ in (\ref{unreducedh}) 
as the {\it kernel} of the Hamiltonian.
We emphasize that the general quartic
form (\ref{unreducedh}) 
includes the local quartic example in (\ref{quarticscalarham}) and 
(\ref{quarticscalarham2}) as well 
as many light-cone models which are non-local in $x^-$ (use Eq.~(\ref{9})). 
The non-local models, including for example the number-conserving 
truncations of local models, can be considered as quantum-mechanical 
models in their own right. 

Following the standard procedure in large $N$ matrix mechanics, we compute
the time derivative of  the annihilation operators
\begin{eqnarray}
\left({{\dot a}_m\over\sqrt{N}}\right)_{rs}&=&
i\left[H.,\left({{a}_m\over\sqrt{N}}\right)_{rs}\right]
=\left(G_m+{\cal E}_m+{\cal F}_m\right)_{rs},\\
\left(G_m\right)_{rs}&\equiv& 
-i \sum_n\left({A_{mn}{a}_n\over\sqrt{N}}\right)_{rs}\\
\left({\cal E}_m\right)_{rs}&\equiv& -i\sum_{p,q}
\left({a^\dagger_p\over\sqrt{N}}\right)_{tu}
\left[\left({a_m\over\sqrt{N}}\right)_{rs}, (A_{pq})_{uv}\right]
\left({a_q\over\sqrt{N}}\right)_{vt} \\
\left({\cal F}_m\right)_{rs}&\equiv& -i\delta_{rs} B_{mq}{1\over N}
\Tr\left({a_q\over\sqrt{N}}\right).
\end{eqnarray}
The term called ${\cal F}_m$ comes from ``opening'' a trace in the double trace
term of the Hamiltonian. To understand this term, we need the 
large $N$ trace theorem
\cite{halperns,halperns81}
\begin{eqnarray}
{1\over N}\Tr f\left({a^\dagger\over\sqrt{N}},{a\over\sqrt{N}}\right)
\ \equalatn\  \bra{0}f\left({a^\dagger},{a}\right)\ket{0}, \qquad \forall f
\label{tracethm1}
\end{eqnarray}
which gives the leading term of the trace at large $N$ in terms of the
reduced average on the right. 
This implies in particular that
\begin{eqnarray}
{1\over N}\Tr \left(a^\dagger_m f_m\left({a^\dagger\over\sqrt{N}},
{a\over\sqrt{N}}\right)\right)
\equalatn {1\over N}\Tr\left( g_m\left({a^\dagger\over\sqrt{N}},
{a\over\sqrt{N}}\right)a_m\right)=0
\label{tracethm2}
\end{eqnarray}
and so the term ${\cal F}_m$ fails to contribute to the large $N$ equation of
motion.

The term called ${\cal E}_m$ collects the contribution of all creation
operators except the first  in the $a^\dagger A a$ term of the Hamiltonian. By
direct computation, we find the explicit form of this term
\begin{eqnarray}
i\left({\cal E}_m\right)_{rs}&=& 
{1\over N}\left[C_{pqm}\left({a^\dagger_p\over\sqrt{N}}\right)_{ts}
\left({a_q\over\sqrt{N}}\right)_{rt}\right.\nonumber\\
&&+C_{pqmx}\left({a^\dagger_p\over\sqrt{N}}\right)_{ts}
\left({a^\dagger_x\over\sqrt{N}}{a_q\over\sqrt{N}}\right)_{rt}
+C_{pqxm}\left({a^\dagger_p\over\sqrt{N}}{a^\dagger_x\over\sqrt{N}}\right)_{ts}
\left({a_q\over\sqrt{N}}\right)_{rt}\nonumber\\
&&\left.+D_{pqmx}\left({a^\dagger_p\over\sqrt{N}}\right)_{ts}
\left({a_x\over\sqrt{N}}{a_q\over\sqrt{N}}\right)_{rt}
+E_{pqxm}\left({a^\dagger_p\over\sqrt{N}}{a_x\over\sqrt{N}}\right)_{ts}
\left({a_q\over\sqrt{N}}\right)_{rt}\right].\label{5*}
\end{eqnarray}
Each of these terms has the form 
$$V_{rs}=\sum_t E_{ts}F_{rt},$$
 called
a ``twisted density'' in Ref.\cite{halperns}.  Twisted densities
can be reduced directly in terms of the tilde operators in 
Eq.~(\ref{tildeops}),
but we prefer to express the result first in terms of untwisted
densities $(ABCD)_{rs}$. Thus we commute the factors into the
form $F_{rt}E_{ts}\to(FE)_{rs}$, keeping all correction terms.

The result of this algebra is
\begin{eqnarray}
i\left({\cal E}_m\right)_{rs}&=& 
{1\over N}\left[\left(C_{pqm}\left({a_q\over\sqrt{N}}{a^\dagger_p\over\sqrt{N}}\right)_{rs}-C_{ppm}\delta_{rs}\right)\right.\nonumber\\
&&+\left(C_{pqmx}\left({a^\dagger_x\over\sqrt{N}}
{a_q\over\sqrt{N}}{a^\dagger_p\over\sqrt{N}}\right)_{rs}-C_{ppmx}
\left({a^\dagger_x\over\sqrt{N}}\right)_{rs}\right)\nonumber\\
&&+\left(C_{pqxm}\left({a_q\over\sqrt{N}}{a^\dagger_p\over\sqrt{N}}{a^\dagger_x\over\sqrt{N}}\right)_{rs}
-C_{ppxm}\left({a^\dagger_x\over\sqrt{N}}\right)_{rs}
-C_{pqqm}\delta_{rs}{1\over N}\Tr{a^\dagger_p\over\sqrt{N}}\right)\nonumber\\
&&+\left(D_{pqmx}\left({a_x\over\sqrt{N}}{a_q\over\sqrt{N}}
{a^\dagger_p\over\sqrt{N}}\right)_{rs}
-D_{pqmp}\delta_{rs}{1\over N}\Tr{a_q\over\sqrt{N}}-D_{ppmx}
\left({a_x\over\sqrt{N}}\right)_{rs}\right)\nonumber\\
&&\left.+\left(E_{pqxm}\left({a_q\over\sqrt{N}}
{a^\dagger_p\over\sqrt{N}}{a_x\over\sqrt{N}}\right)_{ts}
-E_{ppqm}\left({a_q\over\sqrt{N}}\right)_{rs}\right)\right]
\label{6*}
\end{eqnarray}
where each term in parentheses corresponds to one of the
twisted densities in Eq.(\ref{5*}). The two trace terms in
(\ref{6*}) can again be neglected by the large $N$ trace theorem
(\ref{tracethm2}).

    Now all surviving terms in the large $N$ equation of motion are
densities, so we may use reduced matrix elements to  obtain the reduced
equation of motion
\begin{eqnarray}
i{\dot a}_m&=&\left[ H, a_m\right]=A_{mn}(a^\dagger, a)a_n
+{\rm Extra}\phantom{\int}\\
{\rm Extra}&=&
\left(C_{pqm}{a_q}{a^\dagger_p}-C_{ppm}\right)+\left(C_{pqmx}{a^\dagger_x}
{a_q}{a^\dagger_p}-C_{ppmx}{a^\dagger_x}\right)+\left(C_{pqxm}{a_q}{a^\dagger_p}{a^\dagger_x}
-C_{ppxm}{a^\dagger_x}\right)\nonumber\\
&&+\left(D_{pqmx}{a_x}{a_q}
{a^\dagger_p}-D_{ppmx}{a_x}\right)+\left(E_{pqxm}{a_q}
{a^\dagger_p}{a_x}
-E_{ppqm}{a_q}\right)
\end{eqnarray}
where $A_{mn}(a^\dagger,a)$ is the reduced form of the
kernel, given explicitly below.
The parentheses here correspond to the parentheses of Eq.(\ref{6*}).
But now we notice that each of the extra terms in parentheses is separately
zero, e.g.  
\begin{eqnarray}
E_{pqxm}a_qa^\dagger_pa_x=E_{ppxm}a_x
\end{eqnarray}
by the Cuntz algebra.

The results of our computation are the simple reduced equations of motion
\begin{eqnarray}
{\dot a}_m&=&i[H,a_m]=-iA_{mn}(a^\dagger, a)a_n\nonumber\\
{\dot a}^\dagger_m&=&i[H,a^\dagger_m]=ia^\dagger_n A_{nm}(a^\dagger, a)
\label{redeom}
\end{eqnarray}
where $H$ is the reduced Hamiltonian. 

For such simple reduced equations of motion, standard methods again suffice
to write down the {\it reduced Hamiltonian} of the large $N$
light-cone system
\begin{eqnarray}
H-E_0&=&\sum_w a_w^{\phantom{w}\dagger}(\sum_{m,n}a_m^\dagger A_{mn}(a^\dagger,a)a_n)a_w
\label{reducedh}\\
&=&\sum_{m,n}a_m^\dagger A_{mn}(a^\dagger,a)a_n
+\sum_pa_p^\dagger(\sum_{m,n}a_m^\dagger A_{mn}(a^\dagger,a)a_n)a_p
+\cdots\\
A_{mn}(a^\dagger,a)
&=&C_{mn}+{C_{mnp}}a^\dagger_p+
{D_{mnp}}a_p+{C_{mnpq}}a^\dagger_p a^\dagger_q
\nonumber\\
&&\hskip1cm+{D_{mnpq}}a^\dagger_p a_q+{E_{mnpq}}a_p a^\dagger_q
+{F_{mnpq}}a_p a_q\\
&&(H-E_0)\ket{0}=\bra{0}(H-E_0)=0
\end{eqnarray}
where the form $A_{mn}(a^\dagger,a)$ is just the kernel we started with
in the unreduced Hamiltonian (\ref{unreducedh}). The reduced
Hamiltonian is hermitian because $A^\dagger_{mn}=A_{nm}$.
The Cuntz algebra also tells us that the $E$ term of $A_{mn}$
can be further simplified
\begin{eqnarray}
{E_{mnpq}}a_p a^\dagger_q={E_{mnpp}}
\end{eqnarray}
so that the kernel $A_{mn}$ can be considered as 
completely normal-ordered.

In the following section we will argue that the results 
(\ref{redeom}) and (\ref{reducedh})
are in fact the forms obtained by the large $N$ reduction of {\it any} 
light-cone system, where the  reduced kernel 
$A_{mn}$ is always defined by the single
trace term of the Hamiltonian. As a  check on this conclusion, we note 
here
that the reduced equations of motion (\ref{redeom}) 
are consistent for {\it all} $A_{mn}$ 
with the ground state conditions (\ref{groundcond}), for example
\begin{eqnarray}
{\dot a}_m\ket{0}&=&-iA_{mn}(a^\dagger, a)a_n\ket{0}=0.
\end{eqnarray}
Moreover the reduced equations of motion are consistent with the Cuntz
algebra (\ref{cuntzalg})
\begin{eqnarray}
{d\over dt}\left({a}_ma^\dagger_n\right)={d\over dt}\left(
\sum_m{a}^\dagger_ma_m\right)=0
\end{eqnarray}
for all $A_{mn}$. Note that the Cuntz algebra itself is necessary 
\begin{eqnarray}
{d\over dt}\left({a}_ma^\dagger_n\right)=i(-A_{mp}(a^\dagger, a)a_pa^\dagger_n
+a_ma^\dagger_pA_{pn}(a^\dagger, a))=0
\end{eqnarray}
to check 
the first of these relations from (\ref{redeom}).

\section{The General Reduced Light-Cone Hamiltonian}
Now we extend this discussion to a general light-cone 
Hamiltonian. First consider a Hamiltonian, not
necessarily of the light-cone type,
required to be a color singlet and 
behaving at large $N$ no worse than $O(N^2)$. To implement
these restrictions we introduce a Fock space, which
can always be done, but in the general non-light-cone
situation the Fock vacuum is not an energy eigenstate.

To count powers of
$N$, we first recall the well known estimate
\begin{eqnarray}
\bra{.0}\Tr F\left({a_m^\dagger\over\sqrt{N}},{a_m\over\sqrt{N}}\right)
\ket{0.}=O(N)
\end{eqnarray}
with $F$ an arbitrary polynomial of its arguments.
This estimate is a simple consequence of the matrix nature of the operators
and Wick's theorem. The fact that the vacuum expectation
value of a single trace gives the maximal power
of $N$ is embodied in the large $N$ factorization theorem
\begin{eqnarray}
\bra{.0}\Tr F_1\Tr F_2\ket{0.}\sim \bra{.0}\Tr F_1\ket{0.}\bra{.0}
\Tr F_2\ket{0.}=O(N^2).
\end{eqnarray}
Thus a generic color singlet Hamiltonian scaling as $N^2$ will have the form
\begin{eqnarray}
H=N^2e_0+N\Tr h_{11} +\Tr h_{21} \Tr h_{22}+\cdots+{1\over N^{k-2}}
\Tr h_{k1}\Tr h_{k2}\cdots\Tr h_{kk} + \cdots
\label{genericham}
\end{eqnarray}
where $e_0$ is a $c$-number and 
each $h$ is a matrix function of $a/\sqrt{N}$ and $a^\dagger/\sqrt{N}$.
Without loss of generality, we may  put the operators in each
trace in a partially normal-ordered form as explained in Section 2.
This is done by always
moving an annihilation operator at the extreme left of a trace
to the extreme right or a creation operator on the extreme
right to the extreme left. The commutator terms have 
a creation operator and an annihilation operator removed
and an additional trace times $1/N$. They therefore
preserve the generic structure of (\ref{genericham}),
simply redefining the $h_{kl}$. This process is
continued until every trace has the desired form. Note that this
reordering procedure preserves the cyclic ordering within
each trace, but modifies the $h_{kl}$'s in terms with a
smaller number of $a$'s and $a^\dagger$'s.
Moreover, we can assume that each
$h_{kl}$ has zero vacuum expectation value: for if it does not then the VEV
has the right $N$ dependence so that its contribution can
be cancelled by simply
modifying one of the other generic terms. 

Now we describe the further restrictions enjoyed by
light-cone Hamiltonians.
In this case every operator term in the Hamiltonian 
has at least one annihilation
operator and at least one creation operator. In our generic Hamiltonian
(\ref{genericham}) 
we can therefore order things so that in every term there is a creation
operator on the extreme left and an annihilation operator on the
extreme right. With this ordering prescription $\ket{0.}$ is
manifestly an eigenstate of $H$ with eigenvalue $E_0=N^2e_0$. 
Although $\ket{0.}$ is not necessarily the lowest energy eigenstate,
we shall assume in what follows that it is. In more
detail, our ordering prescription shows that the $h$'s have the form
\begin{eqnarray}
 h_{11}= {1\over N}a^\dagger A a; \qquad h_{k1}={a^\dagger\over\sqrt{N}}B_{k1};
 \qquad h_{kk}=C_{kk}{a\over\sqrt{N}}; 
 \qquad h_{kl}={a^\dagger\over\sqrt{N}}B_{kl}+C_{kl}{a\over\sqrt{N}}
\label{lchs}
\end{eqnarray}
for $k>1$ and $1<l<k$. Here we suppress all
indices on the operators. Note that $A$ in $h_{11}$ is the kernel
of the Hamiltonian defined in Sec. 4, and
we remark that Eq.(\ref{genericham}) with the restrictions of (\ref{lchs})
generalizes the discussion of the quartic light-cone field theories
of Sections 2 and 4 to an arbitrary Hamiltonian of the light-cone type.

The first step in inferring the dynamics of the reduced
formulation is to
work out the equations of motion for the unreduced $a_{rs}$'s.
The Heisenberg equation for an annihilation operator is
\begin{eqnarray}
{{\dot a}_{rs}\over\sqrt{N}}
&=&\left(A\left({a^\dagger\over\sqrt{N}},{a\over\sqrt{N}}
\right){a\over\sqrt{N}}\right)_{rs}
+\Tr\ a^\dagger \left[{a_{rs}\over\sqrt{N}}, 
A({a^\dagger\over\sqrt{N}},{a\over\sqrt{N}})\right]a + {\cal F}\\
{\cal F}&=& {1\over N}(\phantom{{a^\dagger\over\sqrt{N}}})_{rs}\Tr()
+{1\over N^2}(\phantom{{a^\dagger\over\sqrt{N}}})_{rs}\Tr()\Tr()+\cdots,
\label{unredeom}
\end{eqnarray}
where we have suppressed the arguments of the various densities
and traces in ${\cal F}$.

The equation of motion for the reduced annihilation operator
should coincide with the unreduced equations of motion
at $N\to\infty$ for matrix elements between the singlet and
dominant adjoint eigenstates as in 
Eqs.~(\ref{wereduction1}) and (\ref{wereduction2}). 
Since the dependence
on the open color indices is completely determined by group
theory, it is sufficient to test the coincidence of equations
of motion in the reduced and unreduced formulations for
these matrix elements with
all the open indices in Eqs.~(\ref{wereduction1}) and (\ref{wereduction2})
saturated.
This is equivalent to testing their coincidence when the
unreduced equation of motion is placed
within single trace VEV's of the form
\begin{eqnarray}
\bra{.0}\Tr(D f(t) E)\ket{0.}, \qquad \forall D, E                         
\end{eqnarray}
where $f(t)=0$ is the unreduced equation of motion
and where $D$ and $E$ are arbitrary matrix functions of the unreduced
$a$'s and $a^\dagger$'s.               
Any simplifications at large $N$ in $f(t)$ valid 
within {\it all} such single
trace VEV's can be exploited in deducing the reduced equations of motion.

We next note the simplifications that occur when the right side 
of (\ref{unredeom}) is inserted in such a VEV. 
 Although every term is nominally of the same order,
the multi-trace terms ${\cal F}$ are all negligible at infinite $N$ because,
by large $N$ factorization,
each trace can be replaced by its
vacuum expectation value, $\Tr(h_{kl})\to\VEV{.0|\Tr(h_{kl})|0.}$.
And the VEV of each $h_{kl}$ vanishes by our
ordering convention. Thus all the terms of ${\cal F}$ in (\ref{unredeom})
can be immediately dropped.

The commutator in the second term on the right side of (\ref{unredeom}) 
is a sum of terms of the form
\begin{eqnarray}
\Tr a^\dagger\left[{a_{rs}\over\sqrt{N}}, 
A({a^\dagger\over\sqrt{N}},{a\over\sqrt{N}})\right]a\sim
{1\over N}(a^\dagger B)_{us}(Ca)_{ru}.
\end{eqnarray}
each of which is a twisted density.  These are the terms which were
most tedious in the example of Sec. 4.
We will show that,  when inserted in the VEV of a single trace, 
such twisted density terms are always suppressed relative
to nominal order by two powers of $1/N$.
Then the right side of the equation of motion
can be replaced by only the first term, and the reduction procedure
is straightforward. 

First we confirm that the first term {\it can} 
contribute at nominal order. So consider
\begin{eqnarray}
\bra{.0}D_{tr}\left(A{a\over\sqrt{N}}\right)_{rs}E_{st}\ket{0.},
\qquad \forall D, E.
\end{eqnarray}
The explicit $a$ to the right of $A$ will contract against an operator in
$E$. If $E$ has a creation operator on the left, 
\ie if $E_{st}=a_{su}^\dagger E_{ut}^\prime/\sqrt{N}$, the contraction
with this operator gives a factor $\delta_{ss}/N=1$ so the resulting
term will give
\begin{eqnarray}
\bra{.0}D_{tr}A_{rs}E^\prime_{st}\ket{0.}=\bra{.0}\Tr(DAE^\prime)\ket{0.}=O(N).
\label{leading}
\end{eqnarray}
Of course, whether this leading term is actually
present depends on the detailed structure of $DAE$, e.g. $\bra{.0}\Tr a^\dagger
a^\dagger a\ket{0.}=0$ even though it is of nominal order $N$.
 
Next we give the corresponding evaluation for the 
twisted density terms on the right side of Eq.~(\ref{unredeom}):
\begin{eqnarray}
\bra{.0}D_{tr} ({a^\dagger\over\sqrt{N}} B)_{us}(C{a\over\sqrt{N}})_{ru}
E_{st}\ket{0.}.                         
\end{eqnarray}               
\newpage
In this case the $a$ on the right will contract only against operators in $E$
and the $a^\dagger$ on the left will contract only against operators
in $D$:
\begin{eqnarray}
\bra{.0}D_{tr} ({a^\dagger\over\sqrt{N}} B)_{us}
(C{a\over\sqrt{N}})_{ru}E_{st}\ket{0.}&=&
\bra{.0}D_{tr} ({a^\dagger\over\sqrt{N}} B)_{us}
{1\over N}(C)_{rv}F_{su}G_{vt}\ket{0.}\nonumber\\
&=&\bra{.0}H_{ta}K_{ur} {1\over{N}}B_{as}
{1\over N}(C)_{rv}F_{su}G_{vt}\ket{0.}.                         
\end{eqnarray}               
Following the color indices, it is easy to see that this
involves a {\it single} trace, albeit in a ``twisted form''. 
But this single trace is multiplied
by the further factor $1/N^2$, so the contribution is nominally of $O(1/N)$.
This is suppressed by two powers of $N$ compared to the
leading contribution (\ref{leading}). We conclude that
\begin{eqnarray}
\bra{.0}\Tr(D \left(i{\dot a(t)}\right) E)\ket{0.} \equalatn 
\bra{.0}\Tr(D \left(A a\right) E)\ket{0.}
\label{veveqmotion}                         
\end{eqnarray}
for $N\to\infty$ for arbitrary matrix operators $D, E$.
In the language of the previous section we have shown that
\begin{eqnarray}
{1\over N}(a^\dagger B)_{us}(Ca)_{ru} &\equalatn& 0\\
i{{\dot a}_{rs}\over\sqrt{N_c}}&\equalatn& 
\left(A\left({a^\dagger\over\sqrt{N}},{a\over\sqrt{N}}
\right){a\over\sqrt{N}}\right)_{rs}
\end{eqnarray}
when sandwiched between the ground state and the dominant
adjoint eigenstates at large $N$.
Since this simple large $N$ equation of motion involves only densities,
it is easily translated into the reduced equation of motion
given in (\ref{redeom})
\begin{eqnarray}
i{{\dot a}}&=& A\left({a^\dagger},{a}\right){a}
\end{eqnarray}
and one finds that the reduced Hamiltonian of the system is that
given in (\ref{reducedh}), now for arbitrary reduced kernel $A$.

Finally we must show that this result does not depend on our
choice of partial normal ordering. Indeed, the different
ordering decisions lead to apparently different expressions
for the kernel $A\left({a^\dagger},{a}\right)$. These differences
occur in the terms in $A$ that are not completely
normal ordered. However, as noted in Sec. 4, use of the
Cuntz algebra
shows that these terms immediately collapse to completely normal-ordered
ones and the discrepancies disappear.

This last point can also be seen in the unreduced formulation.
Consider the contribution of a term in $A$ of the form 
$Ba(x)a(y)^\dagger C/N$
to the right side of Eq.~(\ref{veveqmotion}) 
\begin{eqnarray}
&&\hskip-1cm{1\over N}\bra{.0}\Tr(D Ba(x)a^\dagger(y)Ca E)\ket{0.}
=\bra{.0}\Tr(D BC a E)\ket{0.}c(x,y)\nonumber\\
&&\hskip-0.75cm+\sum {1\over N}\bra{.0}[DB]_{rs}
\Tr [a^\dagger(y)C^\prime ] [C^{\prime\prime} a E]_{sr}\ket{0.}
+\sum {1\over N}\bra{.0}[DB]_{rs}
\Tr [a^\dagger(y)C a E^\prime] 
[E^{\prime\prime}]_{sr}\ket{0.}\nonumber\\
&&\hskip-1cm\sim\bra{.0}\Tr(D BC a E)\ket{0.}c(x,y)
+\sum {1\over N}\bra{.0}
\Tr[DBC^{\prime\prime} a E]\ket{0.}\bra{.0}
\Tr [a^\dagger(y)C^\prime ]\ket{0.}\nonumber\\
&&\hskip-0.75cm
+\sum {1\over N}\bra{.0}\Tr[DB
E^{\prime\prime}]\ket{0.}\bra{.0}\Tr [a^\dagger(y)C a E^\prime] 
\ket{0.}=\bra{.0}\Tr(D BC a E)\ket{0.}c(x,y).
\end{eqnarray}               
Here we have used $x,y$ to collectively denote
all the labels carried by $a, a^\dagger$. Then
we defined $c(x,y)$ by $[a_{rs}(x),a^\dagger_{tu}(y)]=
\delta_{st} \delta_{ru}c(x,y)$.
Thus at infinite $N$, wherever we find the combination $a(x)a^\dagger(y)$
in a term in $A$ we can replace it by the $c$-number $c(x,y)$. 
This is a mirror in the unreduced theory of the
action of the Cuntz algebra in the reduced theory:
After this replacement has been made wherever possible, every term in
the simplified $A$ will be in the completely normal-ordered
form $ a^{\dagger n} a^m$. {This} simplified form of $A$,
which we call $A_\infty$, is then unique. In summary, for the purposes
of the large $N$ limit we may replace the unreduced Hamiltonian
$H.$ by
\begin{eqnarray}
H.\equalatn H_\infty\equiv \Tr a^\dagger A_\infty a 
\end{eqnarray}
where $A_\infty$ is the completely normal-ordered kernel defined above.     
\section{The Algebra of All Normal-Ordered Operators}
Having completed the reduction of the general light-cone theory, we 
discuss in this and the following sections  some simple 
algebraic aspects of the reduced
Hamiltonian systems (\ref{reducedh}) for arbitrary normal-ordered kernel A.  
In particular, the
infinite dimensional algebra developed from the Cuntz algebra in this
section is a simpler version of the infinite dimensional algebra 
developed from the interacting Cuntz algebras in Ref.~\cite{halperns99}.

We begin by expressing the Cuntz algebra as an infinite dimensional 
free algebra \cite{halperns99}
\begin{eqnarray}
a_w a^{\phantom{w^\prime}\dagger}_{w^\prime}&=&
\delta_{w,w^\prime}+\sum_{u\neq0}
(\delta_{w,uw^\prime}a_u+\delta_{w^\prime,uw}
a^{\phantom{u}\dagger}_u)\label{cuntz2}\\
a_w a_{w^\prime}&=&a_{ww^\prime},\qquad a_w^{\phantom{w}\dagger} 
a^{\phantom{w^\prime}\dagger}_{w^\prime}
=a_{w^\prime w}^{\phantom{w^\prime w}\dagger}\label{cuntz1}\\
\delta_{w,w^\prime}&=&\cases{1& when $w=w^\prime$\cr
0 & otherwise\cr}\label{deltacuntz}
\end{eqnarray}
where $w,w^\prime$ and $u$ are arbitrary words.

Next we introduce the general normal-ordered product of Cuntz operators
\begin{eqnarray}
E_{w;w^\prime}\equiv a^{\phantom{w}\dagger}_{w}a_{w^\prime}, \qquad 
E_{w;w^\prime}^{\phantom{w;w^\prime}\dagger}=E_{w^\prime;w},
\quad\forall w, w^\prime
\end{eqnarray}
and we use (\ref{cuntz2},\ref{cuntz1})
to obtain the infinite dimensional algebra of all normal-ordered
Cuntz operators:
\begin{eqnarray}
E_{u;u^\prime}E_{v;v^\prime}&=&\delta_{u^\prime,v} E_{u;v^\prime} 
+\sum_{w\neq0}( \delta_{u^\prime,wv}E_{u;wv^\prime}+
\delta_{v,wu^\prime}E_{wu;v^\prime})\nonumber\\
&=&\sum_{x,x^\prime}L_{uu^\prime;vv^\prime;xx^\prime}E_{x;x^\prime}\label{ee}\\
L_{uu^\prime;vv^\prime;xx^\prime}&=&\delta_{u^\prime,v}
\delta_{u,x}\delta_{v^\prime,x^\prime}
+\sum_{w\neq0}(\delta_{u^\prime,wv}
\delta_{u,x}\delta_{wv^\prime,x^\prime}+\delta_{v,wu^\prime}
\delta_{wu,x}\delta_{v^\prime,x^\prime})\\
\left[E_{u;u^\prime},E_{v;v^\prime}\right]
&=&\sum_{x,x^\prime}
F_{uu^\prime;vv^\prime;xx^\prime}E_{x;x^\prime}\label{algnoo},\qquad
F_{uu^\prime;vv^\prime;xx^\prime}=L_{uu^\prime;vv^\prime;xx^\prime}
-L_{vv^\prime;uu^\prime;xx^\prime}.
\end{eqnarray}
Here $x$ and $x^\prime$ are also arbitrary words.
In spite of appearances, the right sides of
the product relations (\ref{cuntz2}) or (\ref{ee})
yield exactly one term, for
example
\begin{eqnarray}
E_{u;u^\prime}E_{v;v^\prime}&=&E_{u;wx^\prime}\qquad {\rm when}\quad
u^\prime=wv, w\neq0.
\end{eqnarray}
We also note that
\begin{eqnarray}
\left[E_{0;w},E_{w^\prime;0}\right]&=&\delta_{w,w^\prime}-E_{w^\prime;w}
+\sum_{w^{\prime\prime}\neq0}(\delta_{w,w^{\prime\prime}
w^\prime}E_{0;w^{\prime\prime}}+\delta_{w^\prime,w^{\prime\prime}w}
E_{w^{\prime\prime};0})
\end{eqnarray}
where $\delta_{w,w^\prime}$ is a central term proportional to
$E_{0;0}=1$ which commutes
with all the generators of the infinite dimensional algebra.  
Subalgebras of the algebra of normal-ordered
operators are considered in the following section.

Our reduced number and momentum operators ${\hat N}$ and $P$ are easily
written in terms of  these operators
\begin{eqnarray}
{\hat N}&=&
\sum_w a_w^{\phantom{w}\dagger}\left(\sum_m a_m^\dagger a_m\right) a_w
=\sum_{mw}E_{mw;mw}\\
P&=&\sum_w a_w^{\phantom{w}\dagger}\left(\sum_m ma_m^\dagger a_m\right) a_w
=\sum_{mw}mE_{mw;mw}.
\end{eqnarray}
Moreover for a general normal-ordered kernel
$A_{mn}$ we may also express the general reduced Hamiltonian $H^\dagger=H$
as
\begin{eqnarray}
A_{mn}&=&\sum_{u,v}C_{mnuv} E_{u;v}\\
C^*_{mnuv}&=&C_{nmvu},\qquad A^\dagger_{mn}=A_{nm}\phantom{\sum_{u,v}}\\
H-E_0&=&\sum_w a_w^{\phantom{w}\dagger}
(\sum_{m,n}a_m^\dagger A_{mn}(a^\dagger,a)a_n)a_w
=\sum_{w,u,v,m,n}C_{mnuv}E_{umw;vnw}
\label{redham}
\end{eqnarray}
where $u$,$v$ and $w$ are arbitrary words and $C_{mnuv}$ are constants.

Then we may reconsider various commutators above as consequences of the
infinite dimensional
algebra (\ref{algnoo}). For example it is not difficult to check that
\begin{eqnarray}
\left[{\hat N},a_m\right]&=&\sum_{nw}\left[E_{nw;nw},E_{0;m}\right]=-E_{0;m}=-a_m
\nonumber\\
\left[{\hat N},a^\dagger_m\right]&=&\sum_{nw}\left[E_{nw;nw},E_{m;0}\right]=E_{m;0}=
a^\dagger_m
\nonumber\\
\left[P,a_m\right]&=&\sum_{nw}n\left[E_{nw;nw},E_{0;m}\right]=-ma_m
\label{pee1}\\
\left[P,a^\dagger_m\right]&=&\sum_{nw}n\left[E_{nw;nw},E_{m;0}\right]=ma^\dagger_m
\label{pee2}\\
\left[H,a_m\right]&=&-A_{mn}a_n,\qquad \left[H,a^\dagger_m\right]
=a_n^\dagger A_{nm}.\nonumber
\end{eqnarray}
From (\ref{pee1},\ref{pee2}) we find also that the reduced $P$
commutes with all zero momentum operators, e.g.
$$\left[P,a^\dagger_ma^\dagger_na^\dagger_pa_{m+n+p}\right]
=\left[P,E_{pnm;m+n+p}\right]=0.$$
The general form of this statement is
\begin{eqnarray}
\left[P,E_{w;w^\prime}\right]&=&(\{w\}-\{w^\prime\})E_{w;w^\prime}\\
\left[P,E_{w;w^\prime}\right]&=&0 \qquad {\rm when}~ \{w\}=\{w^\prime\}
\end{eqnarray}
where $\{w\}$ is the weight of $w$ (see Eq.~(\ref{lengthw})). 
It also follows that
the reduced momentum operator is conserved
$$\left[P,H\right]=0$$
when the kernel $A_{mn}$ and hence $H$ is constructed from zero momentum
operators. Similarly the reduced number operator commutes with all
zero number operators
\begin{eqnarray}
\left[{\hat N},E_{w;w^\prime}\right]=([w]-[w^\prime])E_{w;w^\prime}\\
\left[{\hat N},E_{w;w^\prime}\right]=0 \qquad {\rm when}~ [w]=[w^\prime],
\end{eqnarray}
where $[w]$ is the length of $w$ (see Eq.~(\ref{lengthw})). 
Then the reduced number operator always
commutes with the reduced momentum operator
\begin{eqnarray}
\left[{\hat N},P\right]=0
\end{eqnarray}
and the reduced number operator is conserved $[H,{\hat N}]=0$ when $A_{mn}$
and hence $H$ is composed of zero number operators.
We shall return to number-conserving reduced Hamiltonians in the
following section.

We also consider the action of these operators on a general
basis state
\begin{eqnarray}
\ket{w}=a_w^{\phantom{w}\dagger}\ket{0}=a^\dagger_{m_n}
\cdots a^\dagger_{m_1}\ket{0},
\qquad\VEV{w|w}=1
\label{reducedbasis}
\end{eqnarray}
and for this discussion the following identities are useful
\begin{eqnarray}
E_{u;v}\ket{w}=\sum_{w^\prime}\delta_{w,w^\prime v}\ket{w^\prime u}\\
\sum_{w^\prime,m,w^{\prime\prime}}\delta_{w,w^\prime mw^{\prime\prime}}
=[w]\label{sumdeltaid}\\
\sum_{w^\prime,m,w^{\prime\prime}}m\delta_{w,w^\prime mw^{\prime\prime}}
=\{w\}.
\end{eqnarray}
Here $[w]$ and $\{w\}$ are the length and weight of $w$,
and we find that
\begin{eqnarray}
{\hat N}\ket{w}=[w]\ket{w},\qquad P\ket{w}=\{w\}\ket{w}
\end{eqnarray}
as expected. The action of the reduced Hamiltonian (\ref{redham}) on the
general basis state is
\begin{eqnarray}
(H-E_0)\ket{w}=\sum_{m,n,u,v,w^\prime,w^{\prime\prime}}C_{mnuv}
\delta_{w,w^\prime vnw^{\prime\prime}}\ket{w^\prime umw^{\prime\prime}}.
\end{eqnarray}
Similarly, one may evaluate the general matrix element of 
$H-E_0$
\begin{eqnarray}
\bra{w^\prime}&=&\bra{0}a_{w^\prime}, \qquad \VEV{w^\prime|w}=\delta_{w^\prime,w}
\nonumber\\
\bra{w^\prime}(H-E_0)\ket{w}&=&
\sum_{m,n,u,v,w^{\prime\prime},w^{\prime\prime\prime}}C_{mnuv}
\delta_{w,w^{\prime\prime\prime} vnw^{\prime\prime}}
\delta_{w^\prime,w^{\prime\prime\prime} umw^{\prime\prime}}
\end{eqnarray}
and so on.

As a simple illustration, we
mention the one-matrix cubic interaction:
\begin{eqnarray}
H&=&\sum_{n=1}^\infty a^{\dagger n}Aa^n,\qquad
A=\omega_0+\lambda(a+a^\dagger), \qquad\ket{n}=a^{\dagger n}\ket{0}\\
(H-E_0)\ket{n}&=&n\omega_0\ket{n}+\lambda(n\ket{n+1}+(n-1)\ket{n-1}), 
\quad n>0.
\end{eqnarray}
We solved this simple recursion relation numerically,
but this solution is superceded by a
recent analytic solution given in Ref.~\cite{durhuus}. 

We finally note that, in large $N$ matrix mechanics, the
large $N$ master field \cite{wittenmaster} is identified \cite{halpern812}
as the set of reduced Heisenberg operators in an
energy eigenbasis. For example, we have
\begin{eqnarray}
(H-E_0)\ket{\alpha}&=&E_\alpha\ket{\alpha},\qquad \alpha=(0,A)\\
a_m(t)&=&e^{iHt}a_m(0)e^{-iHt}\\
a_m(t)_{\alpha\beta}&=&e^{i\omega_{\alpha\beta}t}
a_m(0)_{\alpha\beta},\qquad \omega_{\alpha\beta}\equiv E_\alpha-E_\beta\\
{\dot a}(t)_{\alpha\beta}&=&i\omega_{\alpha\beta}a(t)_{\alpha\beta}
=-iA(t)_{\alpha\gamma}a(t)_{\gamma\beta}
\end{eqnarray}
so that the time dependence of the master field is controlled
by the energy differences $\omega_{\alpha\beta}$ 
among the singlet ground state $\ket{0}$ and
the dominant adjoint eigenstates $\ket{A}$.  These are the time-independent
states which appear in the reduced completeness relation
(\ref{reducedcomplete}). We
return to eigenvalue problems in Sections 7 and 9.

\section{Subalgebras and a Class of Integrable Models at Large N}
We begin this section with a partial list of subalgebras of the
infinite dimensional commutator algebra (\ref{algnoo}) of normal-ordered operators.
\begin{enumerate}
\item The product algebra (\ref{ee}), and hence the
commutator algebra (\ref{algnoo}), of all zero momentum operators
$$E_{w,w^\prime}, \qquad \{w\}=\{w^\prime\}$$
are closed subalgebras. Proof: Consider the product of
two such operators, noting that the Kronecker deltas on the
right of (\ref{ee}) always pick out a unique operator --
which is then a zero momentum operator. (Or note that $P$ commutes with 
a product of two zero momentum operators.)
\item The product and commutator algebras of all zero number operators 
$$ E_{w,w^\prime}, \qquad [w]=[w^\prime] $$
are closed subalgebras. The proof follows as above for
zero momentum operators. In this case we list some
examples
\begin{eqnarray}
\left[E_{m;n},E_{p;q}\right]&=&\delta_{np} E_{m;q}-\delta_{m,q} E_{p;n}\label{a}\\
\left[E_{mn;pq},E_{l;r}\right]&=&\delta_{ql} E_{mn;pr}-\delta_{nr} E_{ml;pq}\label{b}\\
\left[E_{mn;pq},E_{lr;st}\right]&=&\delta_{qr}\delta_{pl} E_{mn;st}
-\delta_{nt} \delta_{ms}E_{lr;pq}\label{c}
\end{eqnarray}
with $m,n,p,q=1\ldots K$. In particular we recognize (\ref{a}) as the algebra of
the unitary group $U(K)$,
where $K$ is the number of lattice sites.
\item Cartan subalgebra and subalgebras thereof. These can be useful in
finding integrable systems.
\end{enumerate}

It is not difficult to find one subalgebra of   the
Cartan subalgebra of (\ref{algnoo}). Consider the hermitian 
``single word'' operators
$$ E_{w;w}=a_w^{\phantom{w}\dagger} a_w, 
\qquad E_{w;w}^{\phantom{w,,w}\dagger}=E_{w;w}, \qquad\forall w $$
which are a subset of the zero momentum and zero number operators. These
satisfy the product relation
\begin{eqnarray}
E_{u;u}E_{v;v}&=&\delta_{u,v}E_{u;v}+\sum_{w\neq0}(\delta_{u,wv}E_{u;u}
+\delta_{v,wu}E_{v;v})\label{freeone}
\end{eqnarray}
and hence each single word operator $E_{u;u}$ is a projection operator
\begin{eqnarray}
\left(E_{u;u}\right)^2&=&E_{u;u}.
\label{eproject}
\end{eqnarray}
Moreover the relation among the projection operators
\begin{eqnarray}
E_{u;u}E_{v;v}&=&\cases{E_{u;u}&when $u=wv$\cr
E_{v;v}&when $v=wu$\cr
0&otherwise\cr}
\label{projectrel}
\end{eqnarray}
is an equivalent form of Eq.~(\ref{freeone}). Finally, we
find that all the single word operators commute
\begin{eqnarray}
\left[E_{u;u},E_{v;v}\right]&=&0,\qquad \forall u, v
\label{freetwo}
\end{eqnarray}
because the
right side of (\ref{freeone}) is symmetric under $u\leftrightarrow v$.

This leads us to consider the following large class of momentum and
number-conserving reduced Hamiltonians
\begin{eqnarray}
A_{mn}&=&\delta_{mn}\sum_w\lambda_w E_{w;w}\\
H-E_0&=&\sum_{wmw^\prime}\lambda_w E_{wmw^\prime;wmw^\prime}
=\sum_{wmw^\prime}\lambda_w a^{\phantom{wmw^\prime}
\dagger}_{wmw^\prime} a_{wmw^\prime}
\label{intham}
\end{eqnarray}
where $\{\lambda_w\}$ is a set of arbitrary word-dependent couplings.
It follows from (\ref{freetwo}) that each of these
models possesses an {\it infinite number of conserved
quantities}
\begin{eqnarray}
{d\over dt} E_{w;w}&=&{d\over dt} \left(a^{\phantom{w}\dagger}_{w}
a_w\right)=0, \qquad \forall w
\label{redcom}
\end{eqnarray}
which leads us to expect that each of the Hamiltonians
(\ref{intham}) is in fact an integrable model.

As further evidence that these models are integrable, we have
checked that their eigenvalue problems are diagonal on 
every basis state $\ket{w}$:
\begin{eqnarray}
(H-E_0)\ket{w}&=&E(w)\ket{w},\qquad
E(w)=\sum_{w^\prime}\lambda_{w^\prime}
\sum_{w^{\prime\prime\prime}mw^{\prime\prime}}\delta_{w,w^{\prime\prime\prime}
w^\prime mw^{\prime\prime}}.
\end{eqnarray}
We can be more explicit by choosing a simple word dependence
$\lambda_w=\lambda([w])$ for the couplings and using the 
identities
\begin{eqnarray}
\sum_{w^{\prime},w^{\prime\prime}}\delta_{w,w^{\prime}w^{\prime\prime}
}&=&[w]+1\\
\sum_{w^{\prime},m,w^{\prime\prime}}\delta_{w,w^{\prime}
mw^{\prime\prime}}&=&[w]\\
\sum_{w^{\prime},m,n,w^{\prime\prime}}\delta_{w,w^{\prime}
mnw^{\prime\prime}}&=&\cases{[w]-1&$[w]\geq2$\cr 0& $[w]=0,1$\cr}\\
&\vdots&\nonumber\\
&&\nonumber\\
\sum_{w^{\prime},w^{\prime\prime}
\atop{\rm letters~of}~u~{\rm at~fixed}~[u]}\delta_{w,w^{\prime}
uw^{\prime\prime}}&=&\cases{[w]-[u]+1 & for $[w]\geq[u]$\cr
0 &for $[w]=0,1,\ldots, [u]-1$\cr}
\end{eqnarray}
which generalize Eq.~(\ref{sumdeltaid}). In this case we find the
eigenvalues
\begin{eqnarray}
E(w)=\sum_{[u]=0}^{[w]-1}\lambda([u])([w]-[u]).
\end{eqnarray}
Note that each term $\lambda([u])$ shows a shifted 
oscillator spectrum $[w]-[u]$.

This class of integrable models is very special.  As 
number-conserving interactions, they correspond to unreduced 
Hamiltonians which are non-local
in $x^-$, but  moreover these particular models provide only a subset
of the number-conserving terms of physical light-cone models
\begin{eqnarray}
\left.\sum_{m,n}{1\over\sqrt{mnp(m+n-p)}}a^\dagger_ma^\dagger_n
a_pa_{m+n-p}\right|_{p=n}=
\sum_{m,n}{1\over {mn}}a^\dagger_ma^\dagger_n
a_na_{m}=\sum_{m,n}{1\over {mn}}E_{nm;nm}.
\end{eqnarray}
We emphasize however that other integrable models would follow 
should the Cartan subalgebra of 
the infinite dimensional algebra (\ref{algnoo})
prove to be larger than the subalgebra (\ref{freetwo}).

\section{Hidden Conserved Quantities at Large N}
In Ref.\cite{halperns} it was pointed out that reduced formulations show
new hidden conserved quantities at large $N$ and, moreover,
that these can be pulled back to find hidden
conserved densities at large $N$ in the unreduced theories.
These features are shared by our reduced light-cone models.

In the first place, the Cuntz algebra itself provides us
with a set of new conserved quantities
\begin{eqnarray}
{d\over dt}\left(a_ma^\dagger_n\right)=
{d\over dt}\left(\sum_ma^\dagger_ma_m\right)=0,\qquad m,n=1\ldots K
\end{eqnarray}
for {\it all} models in our class.
According to the reduction procedure, these conserved quantities 
can be pulled back to find new hidden
conserved densities in the unreduced theory 
\begin{eqnarray}
{d\over dt}\left({a_ma^\dagger_n\over N}\right)_{rs}\equalatn 0,\qquad
{d\over dt}\left(\sum_m{a^\dagger_ma_m\over N}\right)_{rs}\equalatn 0.
\end{eqnarray}
The notation here means that these densities will be conserved
at large $N$, at least when sandwiched between the dominant
unreduced eigenstates $\ket{0.}$ and $\ket{rs,A}$ which saturate the
traces of the theory.

Furthermore, every time we succeed in finding the explicit form
of a reduced trace class operator, we get another hidden
conserved density in the unreduced theory at large $N$. In our
case we find the new densities
\begin{eqnarray}
{d\over dt}P_{rs}&\equalatn&0,\qquad P_{rs}\equiv
\sum_w N^{-[w]-1}\left(a_w^{\phantom{w}\dagger}\left(\sum_m m a^\dagger_ma_m\right)
a_w\right)_{rs}\\
{d\over dt}H_{rs}&\equalatn&0,\qquad H_{rs}\equiv
\sum_w N^{-[w]-1}\left(a_w^{\phantom{w}\dagger}\sum_{m,n} 
a^\dagger_mA_{mn}\left({a^\dagger\over\sqrt{N}},
{a\over\sqrt{N}}\right)a_n
a_w\right)_{rs}
\end{eqnarray}
which correspond to the explicit forms (\ref{reducedpee}) and (\ref{redham}) 
of the reduced $P$ and $H$.
Here the words are constructed as matrix products from the
matrix creation and annihilation operators
$$\left(a_w\right)_{rs}=\left(a_{m_1}\cdots a_{m_n}\right)_{rs},
\qquad \left(a_w^{\phantom{w}\dagger}\right)_{rs}=
\left(a^\dagger_{m_n}\cdots a^\dagger_{m_1}\right)_{rs}.
$$
Following Ref.\cite{halperns}, we remark that the first terms
of the traces of these densities are proportional
to the dominant parts of the original unreduced
 trace class operators $P.$ and $H.$
\begin{eqnarray}
\Tr P&=&{1\over N}P. +
{1\over N^2}\Tr\sum_p\left(a^\dagger_p\left(\sum_m m a^\dagger_ma_m\right)
a_p\right)+\cdots\\
\Tr H&=&
{1\over N}\Tr\left(\sum_{m,n} a^\dagger_mA_{mn}\left({a^\dagger\over\sqrt{N}},
{a\over\sqrt{N}}\right)a_n\right) + \cdots
\end{eqnarray}
but there are an infinite number of other terms from the 
``dressing'' of these densities.

Finally, for each of the integrable models of Sec. 7, we find that the
entire infinite set (\ref{redcom}) of reduced constants of the motion  
can be pulled back
\begin{eqnarray}
{d\over dt}\left(N^{-[w]}a_w^{\phantom{w}\dagger}a_w\right)_{rs}\equalatn 0,
\qquad \forall w
\end{eqnarray}
to appear as hidden conserved densities in the unreduced
theory at large $N$.

\section{Extensions}
\subsection{The dominant states in the unreduced formulation}
Although we have reformulated the large $N$ eigenvalue 
problem in a simpler free-algebraic form, it is also
useful to identify the basis of the dominant adjoint eigenstates
$\ket{rs, A}$ in the unreduced theory. To do this we return
to the discussion of Sec. 5.

In particular, we can insert the unreduced Hamiltonian itself deep inside
the VEV of a trace. By large $N$ factorization the leading behavior
would be
\begin{eqnarray}
\bra{.0}\Tr(D H. E)\ket{0.}\sim\bra{.0}\Tr(DE)\ket{0.}
\bra{.0}H.\ket{0.}=O(N^4)                       
\end{eqnarray}               
since the Hamiltonian is itself a trace. The interesting
physics, however, is contained in the connected part of the VEV which
is of order $N^2$. We can isolate the connected
part by performing the contractions of an
explicit annihilation operator in $H.$ with operators in $E$
or those of an explicit creation operator in $H.$ with  operators
in $D$. Again all the terms in $H.$ with two or
more traces (see Eq.~(\ref{genericham})) are negligible and we obtain
\begin{eqnarray}
\bra{.0}\Tr(D H. E)\ket{0.}-\bra{.0}\Tr(DE)\ket{0.}
\bra{.0}H.\ket{0.}\sim \bra{.0}
\Tr\left(D \Tr\left(a^\dagger A a\right) E\right)\ket{0.}_c
\end{eqnarray}
where the subscript $c$ indicates that at least one of the contractions
noted above is taken. We are thus led to consider the 
following action of $H.$ on the special basis states                     
\begin{eqnarray}
H. \left[\sqrt{N}\left({a^\dagger\over\sqrt{N}}\right)^n\right]_{rs}\ket{0.}
\label{138}
\end{eqnarray}
as was done for singlet states in \cite{thornfock}, where it is shown 
that nearest neighbor
contractions dominate at infinite $N$. The
extra $\sqrt{N}$ in (\ref{138}) normalizes these 
basis states. This is an alternative
path to the Cuntz form of the reduced Hamiltonian at large $N$. 
It is also noteworthy that
the content of the Cuntz algebra in the reduced space is 
essentially summarized by this nearest neighbor interaction
rule. The special basis states in (\ref{138}) are in
1-1 correspondence with the reduced basis states (\ref{reducedbasis}),
and the dominant adjoint eigenstates $\ket{rs,A}$ are linear combinations
of the basis states in (\ref{138}).

\subsection{Energy eigenvalue problem in the singlet sector}
Although we do not have a free-algebraic formulation of
the singlet sector of our matrix models, similar
simplifications have been found in the corresponding unreduced large $N$
singlet eigenvalue problems \cite{thornfock}. For example, we find
in the case of the integrable models of Sec.~7:
\begin{eqnarray}
H.(u)&\equiv& N^{1-[u]}\Tr\left(a^{\phantom{u}\dagger}_u a_u\right)\\
\ket{w.}&\equiv& N^{-[w]/2}\Tr\left(a_w^{\phantom{w}\dagger}\right)\ket{0.},
\qquad \VEV{.w|w.}\equalatn 1\\
H.(u)\ket{w.}&\equalatn& E_u(w)\ket{w.}, \qquad E(u)=\sum_{w\sim vuv^\prime}
\end{eqnarray} 
where $w$ and $u$ are arbitrary words. Multiple trace terms which
arise in this computation are negligible at large $N$ \cite{thornfock}.
The notation used in the eigenvalue $E_u(w)$ was defined in 
Ref.\cite{halperns}: Sum over all words $v$ and $v^\prime$ such that
$vuv^\prime$ is equivalent to $w$ up to cyclic permutation of their
letters.

\subsection{Light-cone Hamiltonian for Non-Abelian Gauge Theory}
When the methods of this paper are applied to Yang-Mills theories,
the unreduced Hamiltonian is much more complex. 
In light-cone gauge, $A_-=0$ the physical potentials are
\begin{eqnarray}
A_i({\bf x},x^-)&=&\int_0^\infty {dp^+\over\sqrt{4\pi p^+}}
\left(a_i(p^+,{\bf x})e^{-ix^-p^+} +a_i^\dagger(p^+,{\bf x})e^{+ix^-p^+}
\right)\\
&\to&\sum_{l=1}^\infty {1\over\sqrt{4\pi l}}
\left(a_{i,l}({\bf x})e^{-ix^-m\delta} 
+a_{i,l}^\dagger({\bf x}) e^{+ix^-m\delta}
\right)
\end{eqnarray}
where we have discretized $p^+$ in the last line. We see that
the operators carry a polarization index $i=1,2$ as well as
a transverse coordinate ${\bf x}$, which can
also be discretized. If one formally eliminates the
longitudinal ``Coulomb'' potential $A_+$, the unreduced Hamiltonian
takes the form
\begin{eqnarray}
H. &=&\int d{\bf x}dx^-{1\over2}\Tr \partial_i A_j\partial_i A_j
+ig\Tr\partial_i A_j[A_i, A_j] +ig\Tr\partial_iA_i{1\over\partial_-}
[A_j,\partial_-A_j]\nonumber\\
&&-{g^2\over2}\Tr A_i A_j[A_i, A_j]
-{g^2\over2}\Tr\left({1\over\partial_-}[A_i,\partial_- A_i]\right)^2\\
&=&\int d{\bf x}dx^-{1\over2}\Tr \partial_i A_j\partial_i A_j
+2ig\Tr A_j{1\over\partial_-}[\partial_-A_i, \partial_jA_i] \nonumber\\&&
\hskip1in-{g^2\over2}\Tr A_i A_j[A_i, A_j]
-{g^2\over2}\Tr\left({1\over\partial_-}[A_i,\partial_- A_i]\right)^2.
\end{eqnarray}
Here the precise definition of the inverse derivatives $\partial_-^{-1}$
has been left unspecified.

The terms arising from partial normal ordering will all be of the
form of a quadratically divergent gluon mass term. There must
be counter terms that  cancel not only this mass, but
also contributions to the mass that arise in higher orders
of perturbation theory. We must also allow for counter terms
which remove potential Lorentz invariance violations induced by ultraviolet
divergences in this noncovariant gauge. Such counter terms need not be 
local in $x^-$.
As a measure of the complexity of this Hamiltonian,
we remark that in 4 space-time dimensions there are 12 cubic terms 
corresponding to different spin transitions, as shown for 
example in \cite{beringrt}. There are clearly many technical
issues that must be resolved before taking this 
Hamiltonian completely seriously.

Even so, it is clear that once $H.$ is completely specified,
the results of this paper show that the reduced
Hamiltonian has the Cuntz-algebraic form discussed above. Burying
all of the unresolved issues in the kernel $A$, we obtain a reduced
Cuntz-algebraic system of the form
\begin{eqnarray}
H-E_0&=&\sum_wa_w^{\phantom{w}\dagger}
\left(\sum_{im{\bf x}jn{\bf y}}a_{i,m}^{\phantom{}\dagger}({\bf x})
A_{im,jn}({\bf x},{\bf y})a_{j,n}({\bf y})\right)a_w\\
a_{i,m}({\bf x})a^\dagger_{j,n}({\bf y})&=&\delta_{ij}\delta_{mn}
\delta_{{\bf x},{\bf y}},\qquad
\sum_{jm{\bf x}}a^\dagger_{j,m}({\bf x})a_{j,m}({\bf x})=1-\ket{0}\bra{0}\\
&&a_{i,m}({\bf x})\ket{0}=\bra{0}a^\dagger_{i,m}({\bf y})=0,\qquad
(H-E_0)\ket{0}=\bra{0}(H-E_0)=0
\end{eqnarray}
where $\ket{0}$ is the reduced ground state.
Moreover, the question of confinement can be addressed by looking
for reduced eigenstates 
\begin{eqnarray}
\ket{{\bf x},{\bf y}}=\sum_{w\in{\cal W}({\bf x},{\bf y})} 
c_wa_w^{\phantom{w}\dagger}\ket{0}
\end{eqnarray}
where ${\cal W}({\bf x},{\bf y})$ is the
set of all words with the transverse
coordinates of the first and last letters constrained
to be ${\bf x}$ and ${\bf y}$ respectively.
For example, if
we could find the lowest energy eigenstate subject to the
constraint ${\bf y}-{\bf x}={\bf R}$, then linear
growth of this energy with $R=|{\bf R}|$ would signal
confinement \cite{rozowskyt}.

\vskip.5cm
\noindent\underline{Acknowledgements.} 
We thank C. Schwartz for many helpful discussions and for
checking the manuscript. This work was supported in
part by the Department of Energy under Grants No. DE-FG02-97ER-41029
and DE-AC03-76SF00098, and in part by the National Science Foundation
Grant PHY-0098840. Also
CBT acknowledges support from the Miller Institute for Basic
Research in Science.


\end{document}